\documentclass[
reprint,%
aps,%
pra,%
groupedaddress,%
superscriptaddress,%
twocolumn,%
assume,%
amsmath,%
floatfix%,
]{revtex4-2}

\usepackage{graphicx}
\usepackage[
  colorlinks=true,
  urlcolor=blue,
  linkcolor=blue,
  citecolor=blue
]{hyperref}
\usepackage{siunitx,braket,amssymb,gensymb,ulem,color}

\newcommand{\rrm}[1]{\textrm{#1}}

\newcommand{\dd}[2]{\frac{\partial#1}{\partial#2}}
\newcommand{\lr}[1]{\left(#1\right)}

\usepackage{graphicx}% Include figure files
\usepackage{dcolumn}% Align table columns on decimal point
\usepackage{bm}% bold math
\usepackage{ORCIDinREVTeX,shellesc}

\begin{document}

\preprint{APS/123-QED}

\title{Gouy phase-assisted Zeno effect for protecting light structure in random media}  

\author{Nilo Mata-Cervera}
\orcid{0000-0001-8464-5102}
\email{nilo001@e.ntu.edu.sg}
\affiliation{Centre for Disruptive Photonic Technologies, School of Physical and Mathematical Sciences, Nanyang Technological University, Singapore 637371, Singapore}
\author{Anton N. Vetlugin}
\orcid{0000-0002-2480-0462}
\affiliation{Centre for Disruptive Photonic Technologies, School of Physical and Mathematical Sciences, Nanyang Technological University, Singapore 637371, Singapore}
\author{Cesare Soci}
\orcid{0000-0002-0149-9128}
\affiliation{Centre for Disruptive Photonic Technologies, School of Physical and Mathematical Sciences, Nanyang Technological University, Singapore 637371, Singapore}
\affiliation{School of Electrical and Electronic Engineering, Nanyang Technological University, Singapore 639798, Singapore}
\author{Miguel A. Porras}
\orcid{0000-0001-8058-9377}
\email{miguelangel.porras@upm.es}
\affiliation{Complex Systems Group, ETSIME, Universidad Politécnica de Madrid, Ríos Rosas 21, 28003 Madrid, Spain}
\author{Yijie Shen}
\orcid{0000-0002-6700-9902}
\email{yijie.shen@ntu.edu.sg}
\affiliation{Centre for Disruptive Photonic Technologies, School of Physical and Mathematical Sciences, Nanyang Technological University, Singapore 637371, Singapore}
\affiliation{School of Electrical and Electronic Engineering, Nanyang Technological University, Singapore 639798, Singapore}

\begin{abstract}
Identifying physical mechanisms that preserve the integrity of structured waves under propagation through complex environments is one of the central problems in wave physics. Here we show that the purity of orbital angular momentum (OAM) modes can be protected against degradation in random media by leveraging two fundamental features of their own Schr\"odinger Hamiltonian dynamics, namely, Zeno effect ---frequent observations slow down the evolution---, and Gouy phase ---the back-action of the observation. Repeated, OAM-dependent Gouy phase kicks imparted along the disturbing path by simple imaging systems trigger the optical Zeno effect that protects the input OAM mode against mode cross-talk that would broaden the OAM spectrum. Given the universality of the mechanism, the Gouy phase-assisted Zeno effect would protect propagation modes other than those of OAM, and the diverse forms of structured light built with them.
\end{abstract}

\maketitle

\section{Introduction}
The simplest form of wave structuring is spatial localization. Then, its propagation phase gets reduced compared to that of a plane wave, a phenomenon known as the Gouy phase anomaly~\cite{gouy1890propriete,gouy1890propagation}. It was first described in optics, but affects all waves whatever their nature ---quantum, acoustic, or electromagnetic~\cite{holme2002gouy,ducharme2015gouy,hiekkamaki2022observation,hamazaki2006direct}---, although its ultimate origin remains a matter of debate~\cite{subbarao1995topological,boyd1980intuitive,feng2001physical}. Gouy phase causes diverse phenomena such as the rotation of the energy flow vector~\cite{padgett1995poynting} and the polarization state~\cite{chen2025gouy}, subluminal group velocity~\cite{bareza2016subluminal,giovannini2015spatially}, longitudinal synchronization and pattern revivals~\cite{da2020pattern,zhong2021gouy}, to name a few. Gouy phase gradual reduction through a focus differs in different wave models and for particular waves~\cite{forbes2021structured,martelli2010gouy,volyar1999topological,cruz2020laguerre,feng1998gouy}.

For paraxial light beams obeying the free Schr\"odinger equation, Gouy phase follows the renowned inverse tangent law of variation for the fundamental mode of propagation, the Gaussian beam, and multiples of it for higher-order modes such as Hermite-Gauss and Laguerre-Gauss (LG) beams. This remarkable mode-dependence of Gouy phase has been exploited to sort different LG modes of a structured beam~\cite{gu2018gouy}. 

Another fundamental feature of the Schr\"odinger dynamics, more acknowledged in quantum physics, is the short-time parabolic decay of the survival probability of an initial quantum state, a feature that makes possible the Zeno effect (ZE). The evolution of the state can be slowed down or even halted through frequent monitoring of whether the state remains the initial one~\cite{Turing,zeno_misra_1,facchi2000quantum}. 
% Initially proposed by Alan Turin~\cite{Turing} as a quantum paradox, then proved by Misra et al.~\cite{zeno_misra_1}, and later generalized to quantum Zeno dynamics by Facchi et al.~\cite{facchi2000quantum}, the ZE has inspired a variety of quantum Zeno evolutions~\cite{zeno_computing_1,QZD_experimental,barontini2025quantum,sazonov2023quantum}.
The original projective measurements in the ZE are often replaced by other models of measurement, such as interaction with other systems (the measuring apparatus) that break the coherence between the system states. Beyond measurements, it is now recognized  that other frequent interruptions of the Hamiltonian evolution that alter the relative phases between states, such as unitary kicks, also produce ZE. \cite{QZD_experimental,zeno_integrated_photonics,blanchard1993strongly,three_examples_QZD,QZD_review}.
\begin{figure}[b!]
    \centering
    \includegraphics[width=\linewidth]{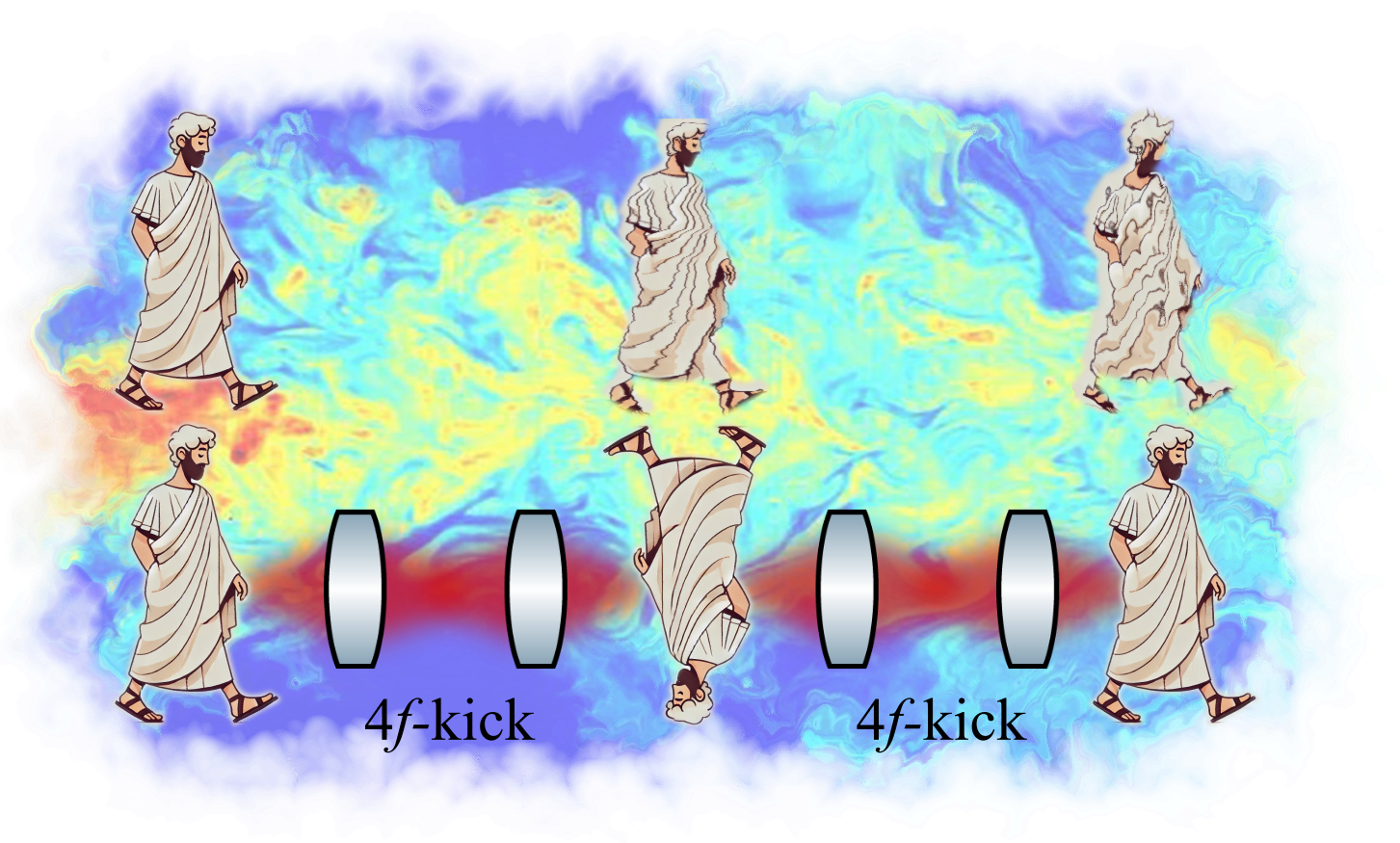}
    \caption{Zeno-protection of light propagating through environment. An image gets distorted when propagates through random medium (top), but preserves its spatial features when it frequently acquires intermodal Gouy phase shifts (bottom) via a $4f-$ system.}
    \label{fig:conceptual}
\end{figure}

\begin{figure*}[t!]
    \centering
    \includegraphics[width=\linewidth]{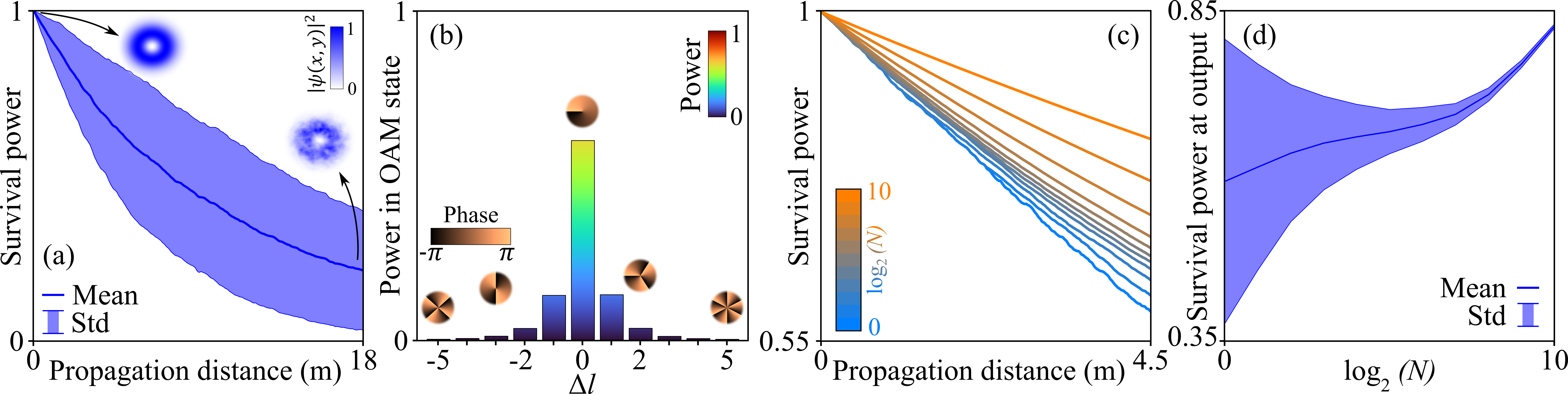}
    \caption{Projective measurement-based Zeno effect in a turbulent medium. (a) In the absence of intermediate measurements, propagation of OAM-carrying light through a turbulent medium of $z_{\rm max}=\SI{18}{\metre}$ results in an exponential decay of the mean survival probability. The inset shows the input $(\ell,p)=(1,0)$ mode and representative output intensity patterns. (b) Averaged over multiple trials OAM spectrum at $z=\SI{4.5}{\metre}$ showing its characteristic broadening. The inset shows the helical phases of the respective subspaces $\Delta\ell$ is the difference between the azimuthal index of the output and input ($\ell=+1$) modes. (c) $N$ projective measurements (equally spaced along the propagation pass) suppress the mode cross-talk resulting in a slower decay of the survival probability. (d) The survival probability at $z=\SI{4.5}{\metre}$ increases and its fluctuation due to the medium randomness decreases with more frequent measurements (larger $N$). Data: $w_0=\SI{3}{\milli\metre}$, $C_n^2=\SI{3e-12}{\meter\tothe{-2/3}}$, $l_0=w_0/20$, $L_0=500\cdot w_0$.}
    \label{f1}
\end{figure*}

The ZE has a classical analog that has been efficiently implemented to modify the evolution of classical light. For example, segmented arrays of coupled waveguides freeze classical tunneling decay from the initial waveguide \cite{longhi_waveguides}, or disconnect tunneling-coupled waveguides~\cite{zeno_integrated_photonics}. Frequent repeaters restoring an image in a communication channel protect it from noise degradation~\cite{gu2012repeater}. In free-space propagation of paraxial light beams, frequent diffracting slits suppress diffraction \cite{diffraction_spreading} and the increasing coherence on propagation \cite{partially_coherent}.

Gouy phase is revealed here as a theoretical tool for Zeno-protection of structured light from degradation in random media, such as atmospheric turbulence~\cite{malik2012influence}. The degradation effect of turbulence on a single input mode of orbital angular momentum (OAM), or LG beam, manifests itself as a mode cross-talk that transfers power to other OAM modes, broadening the OAM spectrum~\cite{structlightturb,klug2021orbital,bachmann2024universal}, and as a decay of the fraction of power in the initial OAM mode ---the survival probability of the initial state in quantum physics. 
In this paper we show that these effects are strongly suppressed when the turbulent propagation is frequently interrupted by Gouy phase kicks. These kicks introduce an OAM-dependent Gouy phase that would be acquired on free space propagation, which in practice are imparted by adequate optics, such as the $4f$ imaging system in the conceptual Fig. \ref{fig:conceptual}. We demonstrate that these repeated Gouy phase kicks produce a ZE that strongly inhibits mode cross-talk and protects the purity of the OAM spectrum.

\section{Light propagation in turbulent media}
Light propagation through complex media is described by the Schr\"odinger equation,
\begin{equation}\label{eq:schrodinger}
    2ik\dd{\psi(\bm{r})}{z}=-\nabla^2_\perp\psi(\bm{r})-2k^2\Delta n(\bm{r})\psi(\bm{r}),
\end{equation}
where the scalar wavefunction $\psi(\bm{r})$ represents the complex envelope of a paraxial field $E(\bm{r})=\psi(\bm{r})e^{i(kz-\omega t)}$ with wavenumber $k=2\pi n_0/\lambda$ and wavelength $\lambda$. The wavefunction $\psi(\bm{r})$ evolves according to the transverse Laplacian $\nabla^2_\perp=\partial^2_{xx}+\partial^2_{yy}$ and the spatially-varying refractive index $\Delta n(\bm{r})=n(\bm{r})-n_0$, with $n_0$ the mean refractive index. 
We use for convenience the transverse coordinate symbol $\bm{r}_\perp=(x,y)$ with $\bm{r}=(\bm{r}_\perp,z)$. The propagator $\hat{P}(z_2, z_1)$ associated with Eq. (\ref{eq:schrodinger}) governs the evolution of light between two planes $z_1$ and $z_2$ of the turbulent medium
\begin{equation}
     \psi(\bm{r}_\perp,z_2) = \hat{P}(z_2, z_1)\psi(\bm{r}_\perp,z_1).
\end{equation}
We first calculate the refractive index $\Delta n(x,y,z)$ within the whole volume of propagation, and then Eq. (\ref{eq:schrodinger}) is solved using a standard split-step numerical algorithm~\cite{turb_mask_1,turb_mask_2}. Each refractive index distribution $\Delta n(x,y,z)$ models a random realization of a turbulent medium described by Kolmogorov's theory~\cite{kolmogorov1991local,imaging_turbulence,turb_mask_2} [see details in Appendix (\ref{sec:A2})]. 
To account for the stochastic nature of turbulence, we model light propagation in multiple ``trials", each time performing an independent random realization of the refractive index volume. Then, we perform statistics of the results by taking mean values and standard deviations (this method is used throughout the paper)~\cite{klug2021orbital,structlightturb}.

As an example, we consider propagation through the turbulent medium of a LG mode $\psi_{\ell,p}(\bm{r})$ with $\lr{\ell,p}=\lr{1,0}$, where $\ell$ and $p$ stand for azimuthal and radial numbers [see  the explicit expression in Appendix (\ref{sec:A1})]. 
Optical beams that propagate in free space as eigenmodes (e.g., LG modes) no longer preserve this property in complex media. 
Indeed, in our example, the $\lr{\ell,p}=\lr{1,0}$ mode, $\psi(\bm{r}_\perp,0)=\psi_{1,0}(\bm{r}_\perp, 0)$, prepared at $z=0$ evolves into a superposition of LG modes, 
\begin{equation}\label{eq:expansion}
\psi(\bm{r})= \sum_{\ell, p}c_{\ell,p}(z)\psi_{\ell,p}(\bm{r}_\perp, z),
\end{equation}
where the propagation-dependent complex coefficients are given by the overlap integral $c_{l,p}(z)=\iint\psi_{l,p}^{\star}(\bm{r}_\perp,z)\psi(\bm{r}_\perp,z)\rrm{d}\bm{r}_\perp$. Equation~(\ref{eq:expansion}) describes mode cross-talk~\cite{bachmann2024universal,peters2025structured} that decreases the survival power (SP),
\begin{equation}\label{eq:survival}
    P(z)=|c_{1,0}(z)|^2,
\end{equation}
of the initial mode.
\begin{figure*}[t!]
    \centering
    \includegraphics[width=\linewidth]{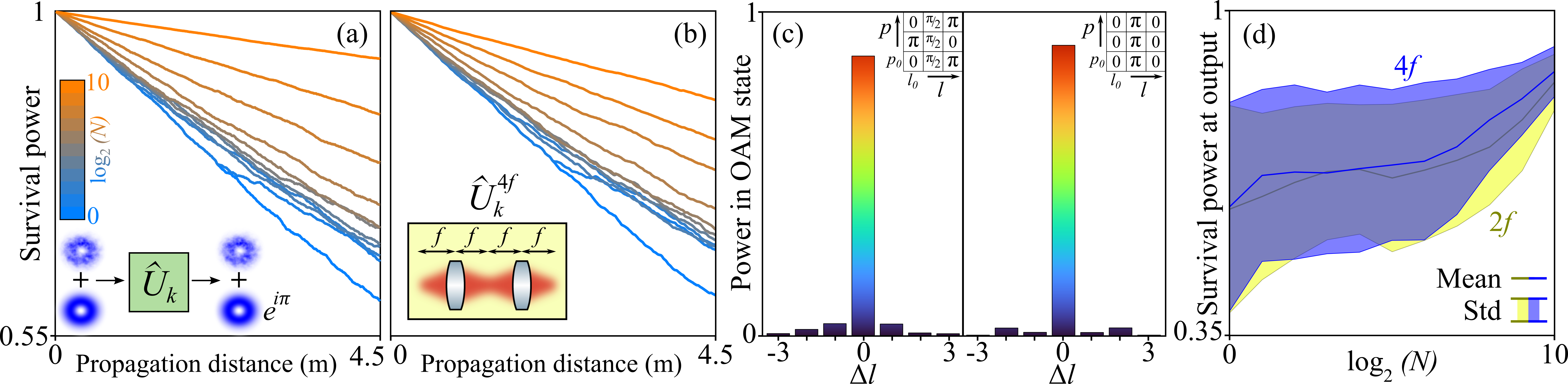}
    \caption{ZE with unitary transformations. Decay of the mean SP for increasing number of (a) $\hat{U}_k$ with $\Phi=\pi$  and (b) $\hat{U}_k^{4f}$ (inset). (c) Averaged OAM spectra at the final plane for $N=2^{10}$ $\hat{U}_k^{2f}$ (left) and $\hat{U}_k^{2f}$ (right) unitary kicks. The inset shows the phase difference $\Delta$ (rad) induced between initial $LG_{\ell_0,p_0}$ and adjacent modes: $\Delta=0$ (green), $\pm\pi/2$ (yellow) and $\pi$ (red). (d) SP at the final plane for increasing number of Gouy-phase kicks. Simulation data is the same as in Fig. \ref{f1}.}
    \label{f3}
\end{figure*}
Figure~\ref{f1}(a) illustrates a characteristic decay of the SP, averaged over multiple trials (thick solid line), accompanied by significant fluctuations due to the randomness of the turbulent medium. An average spectrum of OAM after some propagation in Fig. \ref{f1}(b) reveals a pronounced broadening.

\section{Optical Zeno effect}
\subsection{Projective measurements}
To induce a Zeno effect and suppress transitions to other modes, we can first introduce projective measurements $(\hat{\Pi})$, or OAM-mode filters that retains only the initial mode and eliminate the cross-talk terms in Eq.~(\ref{eq:expansion}):
\begin{equation}\label{eq:measurement}
    \hat{\Pi}\psi(\bm{r}_\perp,z) = c_{l,p}(z) \psi_{l,p}(\bm{r}_\perp,z).
\end{equation}
This operation re-initializes the mode, causing the cross-talk dynamics to restart from the beginning. By introducing $N$ identical, equally spaced projective measurements along the propagation path, 
\begin{equation}
    \hat{\Pi}_N\hat{P}(z_N, z_{N-1})...\hat{\Pi}_2\hat{P}(z_2, z_1)\hat{\Pi}_1\hat{P}(z_1, 0)\psi(\bm{r}_\perp,z),
\end{equation}
the final SP after all measurements is significantly enhanced, as confirmed by modeling in Fig.~\ref{f1}(c). The lower blue curve shows the initial mode's SP decay during uninterrupted propagation, with a single measurement performed at the end ($N=1$). Following Zeno's dichotomy paradox~\cite{aristotle}, the distance between the successive measurements is progressively halved: each subsequent curve represents a modeling of propagation through the same medium, but interrupted by projective measurements every half ($N=2$), quarter ($N=4$), eighth ($N=8$), and so on, of the total propagation distance $z_{\rm max}$. As $N$ increases, mode cross-talk is progressively reduced, leading to higher SP at output, as explicitly shown in Fig. \ref{f1}(d). This is the hallmark of the Zeno effect. In addition, the OAM-carrying light is getting less susceptible to the randomness of turbulent medium as the decreasing standard deviation in Fig. \ref{f1}(d) confirms.

Unfortunately, projective measurements of single LG modes lack practical physical implementation~\cite{gagliardi1976optical,gibson2004free,yan2014high}, are inherently lossy and require a projection onto a single predetermined mode. Nevertheless, the Zeno effect can be achieved by other means, including decoherence, strong coupling and unitary kicks \cite{zeno_integrated_photonics,QZD_experimental,blanchard1993strongly,three_examples_QZD,QZD_review}. Below we show that the ZE driven by unitary kicks can indeed be induced in OAM modes propagating through turbulence by exploiting the mode-dependent nature of the Gouy phase shifts. Furthermore, we show that these unitary kicks could be in principle implemented using elementary optics such as imaging systems.

\subsection{Unitary kicks}
\subsubsection{Mode-selective unitary kicks}
A unitary kick operation, $\hat{U}_k$, introduces a phase shift $\Phi$ between the mode of interest $\psi_{\ell_0,p_0}(\bm{r}_\perp,z)$ and the remaining modes,
\begin{equation}\label{eq:kick}
    \hat{U}_k\psi(\bm{r})= c_{\ell_0,p_0}\psi_{\ell_0,p_0}(\bm{r})+e^{i\Phi}\sum_{\ell,p\neq\ell_0,p_0}c_{\ell,p}(z)\psi_{\ell,p}(\bm{r}),
\end{equation}
with $\Phi=\pi$ producing the strongest ``decoupling" effect~\cite{dhar2006preserving,facchi2004unification,facchi2019kick}. Frequent interruption of propagation through turbulence by unitary kicks reduces the probability of cross-talk of the initial mode with other modes. As shown in Fig. \ref{f3}(a), frequent $\hat{U}_k$ results in an even more pronounced slowdown of the SP decay than with projections. Nonetheless, as happened with projections (\ref{eq:measurement}), we still wonder what optical operation yields $\hat{U}_k$ (\ref{eq:kick}), i.e., attributes different phases to the mode to be protected and to the cross-talk terms.

\subsubsection{Gouy phase-induced unitary kicks}
Among all mode-dependent properties of the LG beams in (\ref{eq:LGbeam}), Gouy phase, $\mathcal{G}_{lp}(z) = -(|\ell|+2p +1)\tan^{-1}(z/z_R)$, where $z_R$ is the Rayleigh distance, is particularly relevant as each mode acquires a different phase on free space propagation. In practice, we do not need free space propagation, but a portion or the complete Gouy phase shift from $\lr{-\infty,\infty}$ can be imparted by simple optics.
For example, when a LG beam passes through a $2f-$focusing system ($\hat{U}_k^{2f}$) with $f=z_R$, it is transformed into 
$\hat{U}_k^{2f}\psi_{\ell p}(\bm{r}_\perp)=\rrm{exp}\lr{i\left[1+|\ell|+2p\right]\pi/2}\psi_{\ell p}(\bm{r}_\perp)$, provided that $\psi_{\ell p}(\bm{r}_\perp)$ is collimated (negligible wavefront curvature). The condition $f=z_R$ just makes the input and output mode components have the same size, thus, it only experiences a phase shift. An arbitrary input beam (\ref{eq:expansion}) in the $2f-$system will be transformed into
\begin{equation}\label{eq:kick2f}
\hat{U}_k^{2f}\psi(\bm{r}_\perp,z)= i\sum_{\ell,p}c_{\ell p}e^{i\lr{|\ell|+2p}\pi/2}\psi_{\ell p}(\bm{r}_\perp,z).
\end{equation}
This is closer to (\ref{eq:kick}), but the kick in (\ref{eq:kick2f}) imparts the same phase to all modes with same $|\ell| + 2p$. The beam shape changes after $\hat{U}_k^{2f}$, since input and output beams are Fourier transform pairs, but each mode only experiences a phase shift. Repeating the similar calculation for a $4f-$system ($\hat{U}_k^{4f}$), this imaging system transforms (\ref{eq:expansion}) into
\begin{equation}\label{eq:kick4f}
\hat{U}_k^{4f}\psi(\bm{r}_\perp,z)=i\sum_{\ell,p}c_{lp}e^{i|\ell|\pi}\psi_{\ell p}(\bm{r}_\perp,z),
\end{equation}
which corresponds to a series of $\pi-$phase shifts between adjacent OAM modes. Unlike $\hat{U}_k^{2f}$, $\hat{U}_k^{4f}$ does not change the shape of the beam, only inverts the image (Fig. \ref{fig:conceptual}), and does not depend on $f$. 

There is a significant difference between projections $\hat{\Pi}_{\rm}$ and kicks $\hat{U}_k$: each individual $\hat{\Pi}_{\rm}$ eliminates all cross-talk terms, effectively ``cleaning" the beam and restarting the evolution. The action of individual $\hat{U}_k$ neither alters the power in each mode, $P_{\ell p}=|c_{\ell p}|^2$, nor the OAM spectrum and its expected value $\langle L_z\rangle=\sum_{\ell p}\ell\cdot|c_{\ell p}|^2$. It is the repeated action of multiple $\hat{U}_k$ that affects the whole evolution, reducing the mode cross-talk spread of the OAM spectrum. Notice that neither $\hat{U}_k^{2f}$ nor $\hat{U}_k^{4f}$ are $\hat{U}_k$ as defined in (\ref{eq:kick}), but they also produce (a less efficient) ZE.

Fig. \ref{f3}(b) depicts the decay of the SP for increasing number of $\hat{U}_k^{4f}$, showcasing a clear ZE. In the same way as frequently monitoring a quantum state prevents its evolution, repeated observations of a light field through optical imaging can protect its OAM states. In each image, no change is done to the beam's structure except for its spatial inversion, and the emerging inter-modal Gouy phase kicks manifest as the back-action of the observation process. The purity of the OAM spectrum after the random medium is also enhanced, as shown in (c) for $\hat{U}_k^{2f}$ (left) and $\hat{U}_k^{4f}$ (right). Although both suppress OAM cross-talk, $\hat{U}_k^{2f}$ only induce $\pi/2$ shifts between adjacent $\ell'$s, while $\hat{U}_k^{4f}$ delays them by $\pi$. Thus, the transition $\Delta\ell=\pm1$ is more probable with $\hat{U}_k^{2f}$ than with $\hat{U}_k^{4f}$ as seen in (c). For $\Delta\ell=\pm2$, $\hat{U}_k^{4f}$ has no effect, but this transition is much more unlikely than $\Delta\ell=\pm1$, so that ZE can still happen. The corresponding SP at $z_{\rm max}$ is shown in (d) as a function of $N$. These simulation results show that mode cross-talk and broadening of OAM spectrum in random media can be reduced significantly just by the back-action of beam focusing or imaging, achieving OAM purities higher than $90\%$ in the present example. All these phenomena as described above are exclusively a consequence of the Gouy phase.

\begin{figure*}[t!]
    \centering
    \includegraphics[width=\linewidth]{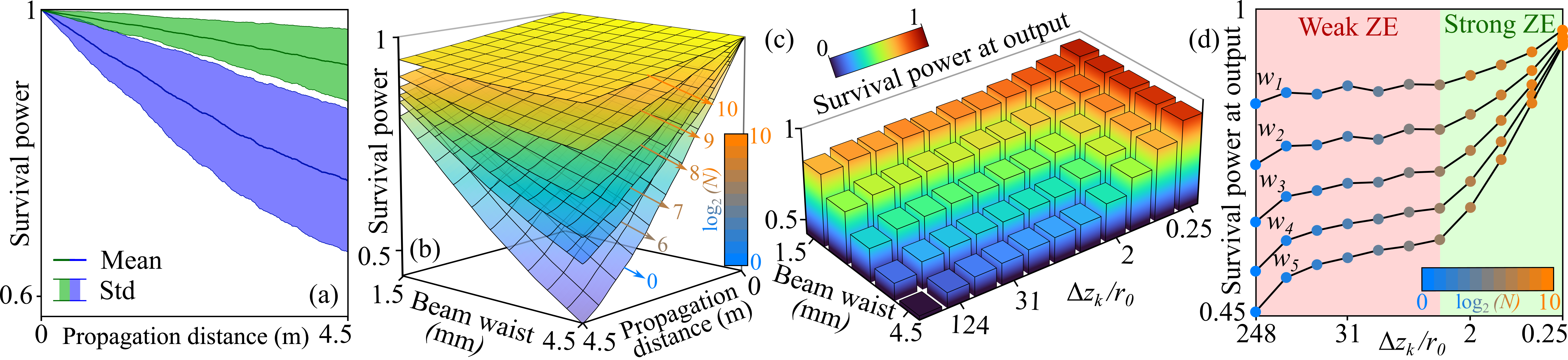}
    \caption{ZE regimes in turbulent media and Zeno distances. (a) Free decay ($N=0$) of the SP for different beam widths $w_0=\SI{1.5}{\milli\metre}$-green and $w_0=\SI{4.5}{\milli\metre}$-blue. (b) Slowing down of the SP decay for increasing number of $\pi-\hat{U}_k$ for different initial widths $w_0$. (c) SP at the final plane for decreasing distance between kicks $\Delta z_k=z_{\rm max}/N$ (increasing kick frequency) and different initial widths $w_0$. (d) Weak and strong Zeno domains for different widths, the transition of which happens when $\Delta z_k$ is of the order of $r_0$. Simulation data is the same as in Fig. \ref{f1} with 400 random realizations.}
    \label{f4}
\end{figure*}

\section{Effect of turbulence strength and Zeno distance}
Finally, we discuss the existence of ZE in stronger turbulent conditions. A key parameter that determines the turbulence strength is the coherence length $r_0$ (\ref{eq:r0}) which measures the scale at which the medium's refractive index becomes uncorrelated~\cite{structlightturb,bachmann2024universal}. The distortion experienced by the beam will be determined by the ratio $\langle r\rangle/r_0$ between the beam size $\langle r\rangle$ and the coherence length, since larger beams experience more randomness than smaller beams in the same medium~\cite{klug2021orbital,bachmann2024universal}. Here we wonder if ZE will still occur if the disturbance experienced by the beam increases, that is, for higher values of $w_0/r_0$.

Beams with larger beam waist $w_0$ present faster SP decays as shown in Fig. \ref{f4}(a). When we introduce an increasing number of $\pi-\hat{U}_k$, the decay of the SP is slowed down, and this phenomenon is observed for any input beam size $w_0$ as shown in (b). It can be seen in previous figures \ref{f1}(b) and \ref{f3}(d) that there is a certain range in which increasing $N$ does not increase significantly the SP at the final plane, but once $N$ exceeds a certain value the power increases much faster. The same behavior is found for $\pi-\hat{U}_k$'s in Fig. \ref{f4}(c), where an abrupt increase of the SP happens once the distance between kicks decreases and becomes comparable with the coherence length $r_0$. This phenomenon is common to all simulated values of $w_0$ but it becomes more obvious when the decay rate is faster, for larger values of $w_0$. As shown in \ref{f4}(d), the larger $w_0$ the lower the SP at the final plane without introducing any $\hat{U}_k$ due to a faster decay rate. At first, the SP increases only very slightly when we increase the frequency of $\hat{U}_k$ - weak ZE. When the distance between $\hat{U}_k$ $(\Delta z_k)$ becomes of the order of the coherence length $r_0$ or smaller the SP increases significantly for larger $N$: strong ZE. This distinguishes two regimes: low interruption frequency $(\Delta z_k\gg r_0$, green shaded area in Fig. ~\ref{f4}(d)) in which turbulence effects dominate, and the Zeno regime $(\Delta z_k\lesssim r_0$, red shaded area in Fig. ~\ref{f4}(d)), in which increasing $N$ reduces mode cross-talk significantly. The transition between these two regimes happens when $\Delta z_k\sim r_0$ for any beam size, suggesting a connection between the coherence length $r_0$ and the Zeno distance -- the distance between measurements that yields ZE, a spatial analog of the Zeno time in quantum systems~\cite{three_examples_QZD,nakazato1996temporal,wilkinson1997experimental,anastopoulos2019decays}. One would expect the Zeno distance to decrease for increasing randomness (higher $C_n^2$), requiring higher interruption rate to maintain a certain OAM purity, which could makes this approach highly unpractical in severe turbulence conditions. On the other hand, if we fix the number of interruptions, their ability to produce Zeno effect will improve as the Zeno distance increases. Given that the results in Fig. (\ref{f1},\ref{f3},\ref{f4}) were obtained with very severe turbulence (the survival power decays within $\SI{4.5}{\metre}$), an additional simulation is carried out scaling the evolution to more realistic distances of the order of hundreds of meters, as shown in the Appendix section (\ref{sec:A3}). In this simulation, the distance between consecutive $\hat{U}_k^{2f}$ increases to the order of tens of centimetres. Nonetheless, even in such moderate turbulence scenario the number of lenses needed to produce an observable Gouy phase-assisted Zeno effect is undesirably large, therefore more practical and efficient decoupling mechanisms may be explored in future research.  
% The feasibility to implement this approach in realistic channels will depend on the turbulence strength: while moderate or weak turbulence requires reasonable distances between interruptions, the required distance for stronger turbulence may become considerably small. 
% This limitation may be overcome by replacing the proposed lens-based Gouy phase kicks by more practical mechanisms. 

\section{Conclusions}
In conclusion, we have demonstrated for the first time the possibility of protecting structured light by the Zeno effect. Specifically, LG modes in random media maintain high mode purity by imparting frequent Gouy phase kicks by simple imaging systems. The universality of the mechanisms involved ---parabolic short-time decay and Gouy phase in the Schr\"odinger wave dynamics--- suggests that the same ingredients could be adapted to protect other forms of structured waves of different nature and in other disturbing environments such as scattering media. The experimental verification of this effect may be carried out by repeatedly impinging a structured beam onto a randomly patterned spatial light modulator (SLM), mimicking the effect of an extended turbulent medium, interrupting the evolution with $2f$ or $4f$ imaging systems followed by analysis of the OAM spectrum with another SLM. Our proposal should also work to protect the purity of the quantum eigenstate of OAM of single photons. In this respect, further research would point towards the protection of entangled photons with OAM by the optical Zeno effect.

\section*{Acknowledgments}
Y.S. acknowledges support from Singapore Ministry of Education (MOE) AcRF Tier 1 grants (RG157/23 \& RT11/23), Singapore Agency for Science, Technology and Research (A*STAR) MTC Individual Research Grants (M24N7c0080), and Nanyang Assistant Professorship Start Up grant. M.A.P. acknowledges support from the Spanish Ministry of Science and Innovation, Gobierno de Espa\~na, under Contract No. PID2021-122711NB-C21. A.N.V. and C.S. acknowledge support from the National Centre for Integrated Photonics (NRF-MSG-NCAIP).
\section*{Data availability}
The simulation data is openly available at this \href{https://doi.org/10.21979/N9/C3FOYV}{link}. 
\section{Appendix}
\subsection{Laguerre-Gaussian modes}\label{sec:A1}
\renewcommand{\theequation}{A\arabic{equation}}
\setcounter{equation}{0}
Through this work, we have used the Laguerre-Gaussian (LG) modes, solution to the paraxial wave equation in cylindrical coordinate system $r=\sqrt{x^2+y^2}$, $\varphi=\tan^{-1}\lr{y/x}$ given by
\begin{widetext}
    \begin{equation}\label{eq:LGbeam}
    \psi_{\ell,p}(r,\varphi,z)=\sqrt{\frac{2p!}{\pi(|\ell|+p)!}}\frac{1}{w(z)}\lr{\frac{\sqrt{2}r}{w(z)}}^{|\ell|}\rrm{exp}\lr{-\frac{r^2}{w^2(z)}}L_p^{|\ell|}\lr{\frac{2r^2}{w^2(z)}}\rrm{exp}\lr{\frac{ikr^2}{2R(z)}+i\ell\varphi+ikz+i\mathcal{G}(z)}
\end{equation}
\end{widetext}
where $\ell$ and $p$ are the azimuthal and radial indices respectively, $w(z)=w_0\sqrt{1+\lr{z/z_R}^2}$ is the beam width with $w_0$ the beam waist, $z_R=kw_0^2/2$ the Rayleigh range and $k=2\pi/\lambda$ the wavenumber for a given wavelength $\lambda$. $\mathcal{G}(z)=-(2p+|\ell|+1)\tan^{-1}{\lr{z/z_R}}$ is the Gouy phase and $R(z)=z[1+\lr{z_R/z}^2]$ is the wavefront radius of curvature.

\subsection{Simulation of Kolmogorov's turbulence}\label{sec:A2}
\setcounter{equation}{0}
\renewcommand{\theequation}{B\arabic{equation}}
The statistics of turbulent media were analyzed by Kolmogorov using structure functions, which describe the mean fluctuations of the thermodynamic variables such as pressure or temperature~\cite{kolmogorov1991local,turb_mask_1,turb_mask_2}. These fluctuations induce spatial and temporal variations of the optical properties of the atmosphere. In particular, the random refractive index of a turbulent atmosphere can be described by a power spectral density given by Von Kármán
\begin{equation}\label{eq:PSD}
    \Phi_n(\kappa)=0.033\cdot C_n^2\cdot\frac{\rrm{exp}\lr{-\kappa^2/\kappa_m}}{\lr{\kappa^2+\kappa_0^2}^{11/6}},
\end{equation}
where $C_n^2$ is the refractive index structure constant, a measure of the strength of the refractive index fluctuations, $\kappa_m=5.92/l_0$, $\kappa_0=2\pi/L_0$ with $l_0$ and $L_0$ the inner and outer scales of turbulence. $\bm{\kappa}$ is the coordinate vector in reciprocal space with $\kappa^2=\kappa_x^2+\kappa_y^2+\kappa_z^2$. In our simulations we are in the inertial sub-range $l_0\ll r\ll L_0$, in which air flow is isotropic and Kolmogorov's theory remains valid. A key magnitude that determines the optical disturbance introduced by turbulence is the Fried parameter. It has dimensions of length, and represents the distance between two points of the medium at which the refractive indices become uncorrelated~\cite{structlightturb}, also known as coherence length
\begin{equation}\label{eq:r0}
    r_0=\lr{0.423k^2\int_0^{z_{\rm max}}C_n^2(z)dz}^{-3/5}=\lr{0.423k^2C_n^2z_{\rm max}}^{-3/5},
\end{equation}
where we considered constant $C_n^2$ in the propagation~\cite{structlightturb}. Here, $k=2\pi/\lambda$ is the wavenumber, $z_{\rm max}$ is the length of the turbulent medium.

The refractive index of turbulence $\Delta n(x,y,z)$ is generated in a volume $L_x\times L_y\times z_{\rm max}$, where $L_x=L_y$ are the transverse dimensions of the simulation window and $z_{\rm max}$ the total propagation distance. To construct it we follow a standard Fourier transform technique~\cite{turb_mask_1,turb_mask_2}, where the refractive index Fourier coefficients $c_{\kappa_x,\kappa_y,\kappa_z}$ are Gaussian random variables with zero mean $\langle c_{\kappa_x,\kappa_y,\kappa_z}\rangle=0$ and variance given by Eq. (\ref{eq:PSD}) $\langle |c_{\kappa_x,\kappa_y,\kappa_z}|^2\rangle=\Phi_n(\kappa)$. Then, the refractive index in real space $\Delta n(x,y,z)$ is obtained by 3D inverse Fourier transform with the addition of $3^3=27$ sub-harmonics to increase the numerical accuracy~\cite{imaging_turbulence}. 

We solve Eq. (\ref{eq:schrodinger}) with a split-step approach~\cite{turb_mask_1,turb_mask_2} by discretizing a refractive index volume into $\mathcal{P}$ discrete propagation planes within the total propagation distance $z_{\rm max}$. Numerically, between two consecutive planes the wave evolves a free-space distance $\delta=z_{\rm max}/\mathcal{P}$ followed by a phase mask extracted from the volume refractive index distribution at that plane~\cite{taha1984analytical}.
\subsection{Dimensional analysis}\label{sec:A3}
\setcounter{equation}{0}
\renewcommand{\theequation}{C\arabic{equation}}
Two different evolutions in turbulent media share the same dynamics when the associated adimensional variables match~\cite{klug2023robust}. The disturbance experienced by the beam is determined by the adimensional variable $r_0/w_0$, where $r_0$ is the coherence length (\ref{eq:r0}) and $w_0$ is the beam width. Here we also impose that two beams experience the same diffraction, which in the LG beam optics implies that the dimensionless distance $z_{\rm max}/z_R$ is the same. Here $z_R$ is the Rayleigh range, and $z_{\rm max}$ is the total propagation distance. Between two evolutions within $z_{\rm max}$ and $z'_{\rm max}$ we get the following equality
$$\frac{z}{z_R}=\frac{z'}{z_R'}\rightarrow \alpha_z=\frac{z'}{z}=\lr{\frac{w_0'}{w_0}}^2=\alpha_{w_0}^2\rightarrow \alpha_{w_0}=\sqrt{\alpha_z}.$$
This fixes the relation between beam waists and total distances in order the two beams to experience the same diffraction. To match the disturbances we require $r_0/w_0=r_0'/w_0'$ so that $\alpha_{r_0}=r_0'/r_0=w_0'/w_0=\alpha_{w_0}.$ Finally, with the definition of the coherence length $r_0=\lr{0.423k^2C_n^2z}^{-3/5}$ for a medium with constant turbulence strength $C_n^2$, the relation between turbulence strengths is given by
$$\frac{C_n^{2'}}{C_n^2}=\alpha_z^{-11/6}.$$
Thus, for a given turbulence strength $C_n^2$ and propagation distance $z_{\rm max}$, a dimensionally equivalent evolution is obtained in a total distance $z'_{\rm max}=z_{\rm max}\cdot\alpha_z$ if the structure constant is scaled as $C_n^{2'}=C_n^2\alpha_z^{-11/6}$ and the beam waist as $w_0'=w_0\sqrt{\alpha_z}$. As an example, below we show the results for $z_{\rm max}=\SI{1000}{\metre}$, dimensionally equivalent to the simulation in Fig.~\ref{f3} with $\alpha_z=1000/4.5\approx222.2$, simulated with the parameters $w_0'=w_0\sqrt{\alpha_z}\approx\SI{4.5}{\centi\metre}$, $C_n^{2'}\approx\SI{1.5e-16}{\metre^{-2/3}}.$
\begin{figure}
    \centering
    \includegraphics[width=0.75\linewidth]{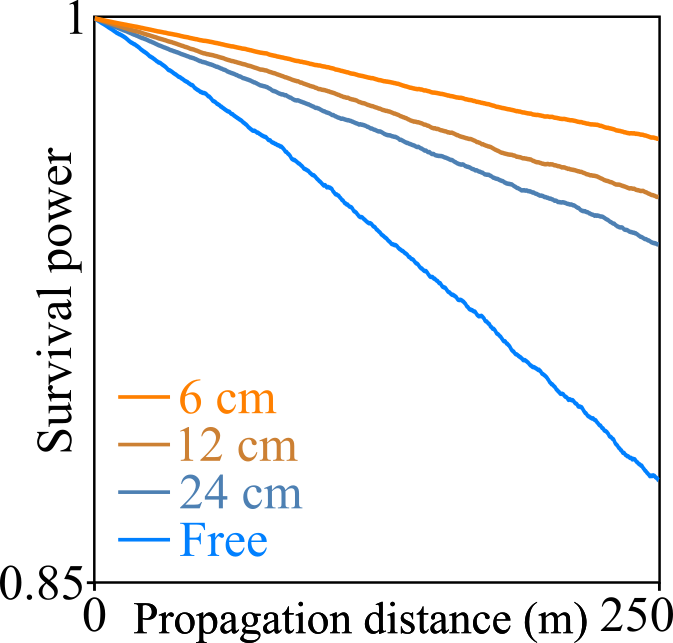}
    \caption{Decay of the survival power for increasing number of imaging systems $\hat{U}_k^{2f}$. The lower curve shows evolution without $\hat{U}_k^{2f}$, and the distance between consecutive $\hat{U}_k^{2f}$ is shown in the inset. Data $C_n^2=\SI{1.5e-16}{\metre^{-2/3}}$, $z_{\rm max}=\SI{250}{\metre}$, $w_0=\SI{4.5}{\centi\metre}$, $\ell=1$, $p=0$. Results averaged over $400$ random realizations.}
    \label{fig:placeholder}
\end{figure}

\sloppy
\bibliography{bibliography}

\end{document}